# Production of $\eta$ Mesons in 200 $A$GeV/$c$ S+S and S+Au Reactions

R. Albrecht[a], V. Antonenko[g], T.C. Awes[e], C. Barlag[d], F. Berger[d], M.A. Bloomer[b], C. Blume[d], D. Bock[d], R. Bock[a], E.M. Bohne[d], D. Bucher[d], A. Claussen[d], G. Clewing[d], R. Debbe[f], L. Dragon[d], A. Eklund[c], S. Fokin[g], S. Garpman[c], R. Glasow[d], H.Å. Gustafsson[c], H.H. Gutbrod[a], O. Hansen[f,1], G. Hölker[d], J. Idh[c], M. Ippolitov[g], P. Jacobs[b], K.H. Kampert[d,2], K. Karadjev[g], B.W. Kolb[a], A. Lebedev[g], H. Löhner[h], I. Lund[h], V. Manko[g], B. Moskowitz[f], S. Nikolaev[g], F.E. Obenshain[e], A. Oskarsson[c], I. Otterlund[c], T. Peitzmann[d], F. Plasil[e], A.M. Poskanzer[b], M. Purschke[a], B. Roters[a], R. Santo[d], H.R. Schmidt[a], K. Söderstrom[c], S.P. Sørensen[e,i], P. Stankus[e], K. Steffens[d], P. Steinhaeuser[a], E. Stenlund[c], D. Stüken[d], A. Vinogradov[g], H. Wegner[f,3], and G.R. Young[e]

**WA80 Collaboration**

a. *Gesellschaft für Schwerionenforschung, D-64220 Darmstadt, Germany*
b. *Lawrence Berkeley Laboratory, Berkeley, California 94720, USA*
c. *University of Lund, S-22362 Lund, Sweden*
d. *University of Münster, D-48149 Münster, Germany*
e. *Oak Ridge National Laboratory, Oak Ridge, Tennessee 37831, USA*
f. *Brookhaven National Laboratory, Upton, New York 11973, USA*
g. *Russian Research Center "Kurchatov Institute", Moscow 123182, Russia*
h. *KVI, University of Groningen, NL-9747 AA Groningen, Netherlands*
i. *University of Tennessee, Knoxville, Tennessee 37996, USA*

**Abstract**

Minimum Bias production cross sections of $\eta$ mesons have been measured in 200 $A$GeV/$c$ S+Au and S+S collisions at the CERN SPS by reconstructing the $\eta \to \gamma\gamma$ decay. The measurements have been made over the rapidity range $2.1 \leq y \leq 2.9$ using the leadglass spectrometer of WA80. Within the statistical and systematical uncertainties the spectral shapes of $\pi^0$ and $\eta$ mesons yields are identical when their invariant differential cross section is plotted as a function of the transverse mass. The relative normalization of the $\eta$ to $\pi^0$ transverse mass spectra is found to be $0.53 \pm 0.07$ for S+Au and $0.43 \pm 0.15$ for S+S reactions. Extrapolation to full phase space leads to an integrated cross section ratio of $\eta$ to $\pi^0$ mesons of $0.147 \pm 0.017$(stat.) $\pm$ 0.015(syst.), and $0.120 \pm 0.034$(stat.) $\pm$ 0.022(syst.) for S+Au and S+S collisions, respectively.

[1] Now at the Niels Bohr Institute, University of Kopenhagen, Denmark
[2] Now at University of Karlsruhe, Germany
[3] deceased

# 1 Introduction

The measurement of inclusive transverse momentum distributions of produced particles and their relative production cross sections is known to be a valuable tool in the study of high energy hadronic interactions. While many efforts have been undertaken to measure charged hadrons in ultra-relativistic nuclear collisions at the CERN-SPS and BNL-AGS [1], up to now only little is known about the production of neutral mesons heavier than the $\pi^0$. The WA80 collaboration at the CERN-SPS has therefore concentrated on identifying $\pi^0$ and $\eta$ mesons with high precision over an extended transverse momentum range via their two-photon decay channel. Being the lightest mesons, with the $\eta$ meson containing (hidden) strangeness, they provide important information about the dynamics and decay of the hot fireball formed in central nucleus-nucleus collisions.

In pp collisions, relations between the production cross sections for different mesons have been discussed within the phenomenological concept of scaling with transverse mass $m_T = \sqrt{m_0^2 + p_T^2}$. In this picture, the shapes of the differential cross sections of mesons, plotted as a function of $m_T$ are expected to be identical. As a simple consequence of different rest masses, this leads to a strong $p_T$-dependence of the cross section ratio at low values of $p_T$ (see e.g. [2, 3] and Fig. 5). For values of $p_T \gg m_0$, on the other hand, the meson yield ratios were found to become constant and independent of the reaction system and CM energy. Some qualitative interpretations of this behavior are discussed in the literature [4, 5, 6].

Previous data on mid-rapidity pion and kaon production in nuclear reactions at CERN [7, 8] and BNL [9] seem to be consistent with this picture. However, the dynamic range in these experiments generally spans only a limited transverse momentum range at low $p_T$ and their statistical and systematic errors still leave some freedom. In the present work we will extend these studies to neutral pions and $\eta$-mesons. A possible scenario, in which a deviation from $m_T$ scaling may be observed, is an expanding fireball including collective transverse flow, because in this picture thermal and flow components affect mesons with different masses differently [10]. Therefore, in the presence of a large collective flow component and similar freeze-out conditions, the $m_T$ spectra of heavier mesons are expected to become flatter. The present absolute cross section measurements of $\pi^0$ and $\eta$ mesons with approximately 3 GeV/$c$ overlap in transverse momentum and with identical rapidity acceptance provide a stringent test of the $m_T$ scaling concept.

# 2 Experimental

The WA80 experiment at the CERN-SPS (Fig. 1) is equipped with a finely segmented leadglass spectrometer covering the polar range $6.2° \leq \vartheta_{\text{lab}} \leq 13.9°$. Compared to preceding WA80 data taking runs, the leadglass calorimeter SAPHIR [11] has been augmented by two additional leadglass calorimeters (TOWERS) in order to achieve an improved and uniform acceptance for $\pi^0$ and $\eta$ mesons in the rapidity range $2.1 \leq y \leq 2.9$. Both meson types are reconstructed by invariant mass analysis of their $\gamma\gamma$ decay mode. The total azimuthal coverage of the 3798 leadglass modules at a distance of 9 m to the target is $\sim 1.3\,\pi$. The SAPHIR modules consist of SF5 leadglass blocks of dimension $3.5 \times 3.5 \times 46$ cm$^3$, whereas the TOWER modules are made of $4.0 \times 4.0 \times 40$ cm$^3$ TF1 leadglass.

The minimum bias trigger requires a valid signal of the beam counters and a minimum amount of transverse energy $E_T \gtrsim 1$ GeV, detected by the Mid-Rapidity Calorimeter [12]. Data have been taken with the 200 $A$GeV/$c$ sulfur beam on targets of Au (250 mg/cm$^2$) and S (205 and 510 mg/cm$^2$). For the present analysis 7.9 million S+Au and 1.4 million S+S minimum bias events were accumulated. The $p_T$ distributions for $\gamma$'s and $\pi^0$'s from S+Au interactions could be determined over a wide range of $0.3 \leq p_T \leq 4.5$ GeV/$c$. For the first time also $\eta$ meson yields are measured in ultrarelativistic heavy ion reactions over the $p_T$ range of $0.5 \leq p_T \leq 3.5$ GeV/$c$.

# 3 Data Analysis

In addition to modifications of the experimental setup more refined analysis methods have been developed to improve the particle identification performance compared to previous WA80 analyses [13, 14]. Most importantly, a more efficient distinction between hadronic and electromagnetic showers is achieved by analyzing their lateral dispersion. Furthermore, an improved procedure for unfolding overlapping showers was used [15]. Different than from other works [16], the charged particle veto was not used in the present analysis.

In order to reconstruct $\pi^0$ and $\eta$ mesons from the experimental data and to extract their differential cross sections, $d\sigma/dp_T$, the invariant mass of each $\gamma\gamma$-pair of an event is calculated and sorted as a function of the transverse momentum of the photon-pair. The resulting invariant mass distributions are composed of peaks centered at values of the $\pi^0$ and $\eta$ rest mass and are located on top of a large combinatorial background. The meson yields are extracted from these distributions by subtracting the combinatorial background. Possible contaminations in the photon sample, e.g. due to non-rejected charged hadrons (see above), thus do only contribute to the uncorrelated (and subtracted) combinatorial background, but not contribute to the isolated invariant mass peaks itself. Due to the high particle multiplicity in heavy-ion collisions the resulting signal-to-background ratios may become very unfavorable particularly at transverse momenta below 1 GeV/$c$, so that precise knowledge of the shape and yield of the combinatorial background below the mass-peaks is required. Instead of fitting the observed background distributions by arbitrarily chosen continuous functions, we have therefore applied an event-mixing procedure to calculate the background distributions from the experimental data itself [17, 18]. Special care has been taken to ensure that the data and background distributions are generated for the same class of events with very similar phase space characteristics. Therefore, photons from a given event class have been combined exclusively with photons from another event of the same centrality class, which have been selected using 8 equidistant bins of the associated transverse energy $E_T$. To minimize the statistical and systematic uncertainties in the extraction of the $\eta$ peak contents, a two-dimensional fit of the data/background distribution as a function of $p_T$ and invariant mass has been performed. Based on the measured response function of the calorimeter, the peak has been assumed to be of Gaussian shape with position and width taken as free parameters but fixed to be the same for all $p_T$ bins.

To validate the procedure of backgound generation and to check the accuracy of the meson reconstruction procedure, a Monte Carlo simulation was performed in which a known number of $\pi^0$ and $\eta$ mesons were generated according to the experimentally observed distributions. After subjecting those photons found within the experimental acceptance to realistic detector resolution effects, the same analysis techniques as used for the experimental data were applied. The number of mesons extracted agrees with the known number of generated particles [19, 18] within the statistical errors of the background subtraction procedure. Using similar statistics as in the experimental data, systematic deviations imposed by the background subtraction were found to be about 2% for the $\pi^0$ and 10% for $\eta$ distributions.

Fig. 2 shows $\gamma\gamma$ invariant mass distributions for the S+Au data in the $\pi^0$ and $\eta$ mass range after background subtraction. Whereas the significance of the $\pi^0$ peak is obvious, that of the $\eta$ peak needs to be proven. All invariant mass distributions were thus fitted in the range of the $\eta$ peak both by assuming a constant (fit a) and the sum of a Gaussian plus a constant (fit b). The ratio of the obtained chi-squares, $r = \chi^2_{\nu,a}/\chi^2_{\nu,b}$, yields the significance of the peak via an $F$-test [20]. In the case of the $\eta$ peak of Fig. 2b, the significance level is approx. 95 % and even the fits obtained for narrower $p_T$ ranges result in significance levels between 70 % and 90 %. The position and width of the experimental peaks were found to be compatible with the known $\eta$ mass of $m_\eta = 547.45$ MeV/c$^2$ and with the expected invariant mass resolution, $\sigma_\eta \cong 25$ MeV/c$^2$, taking into account the energy and position resolution of the leadglass calorimeters [21].

The $\pi^0$ and $\eta$ distributions are corrected for the geometrical acceptance, which considers the loss of photons due to the incomplete coverage of the full solid angle and the software imposed low energy photon cut-off of $E_{\rm min} = 750$ MeV. Furthermore, the reconstruction efficiency has been determined in order to correct for losses during analyis of photons, and thereby $\pi^0$ and $\eta$ mesons, due to the shower reconstruction procedure. Such a correction is of particular importance in high multiplicity heavy-ion reactions where a large probability for overlapping showers is encountered which may alter their reconstructed energies, positions, and transverse dispersions. To calculate the efficiency, GEANT simulated photon showers, which were well adjusted to the measured detector response, were superimposed onto real data events. By analyzing such artificial events with the same package of shower reconstruction programs as used for real data, the reconstruction efficiency for those superimposed particles could be extracted. A detailed description of this method [21, 19] will be presented in a forthcoming publication. Finally, we point out that effects of reconstruction efficiencies and other systematic errors largely cancel in the $\eta/\pi^0$-ratio as a function of $m_T$, because both mesons were identified via the same $\gamma\gamma$-decay mode and have similar energy for fixed transverse mass of the parent particle. The total systematic uncertainty of up to approx. 11 % in the extracted $\eta/\pi^0$ ratios thereby are dominated largely by the extraction of the $\eta$ peak contents from the invariant mass spectra.

# 4 Results and Discussion

Acceptance and efficiency corrected invariant cross sections of $\pi^0$ and $\eta$ mesons are shown in Fig. 3 for minimum bias S+Au and S+S reactions as a function of the transverse mass $m_T$. In addition to the statistical errors, the systematic errors of the acceptance and efficiency correction have been added quadratically. The invariant cross section of $\pi^0$ mesons in both systems clearly exhibit a spectrum with concave shape. A similar shape is observed for the $\eta$ mesons. This so called $m_T$ scaling behavior of the invariant cross sections is supported by a variety of experimental results and has been observed for a wide range of center of mass energies and for different meson masses [7, 9, 22, 23, 24]. Phenomenological interpretations are mostly based on thermal particle production mechanisms and are discussed for example in Refs. [4, 5, 6].

In order to analyze the data in more detail, an empirical power-law formula (Eq. 1) proposed in Ref. [5]

$$E\frac{d^3\sigma}{dp^3} = C\left(\frac{p_0}{p_0 + m_T}\right)^n \tag{1}$$

has been fitted to the $\pi^0$ $m_T$ spectra of Fig. 3 with the free parameters $C$, $p_0$, and $n$. For the S+Au data the extracted values are $p_0 = 6.8 \pm 0.1$ GeV/$c$ and $n = 37.5 \pm 0.3$. The results are displayed in Fig. 3 as solid lines. The dotted lines, which represent fits to the $\eta$ distributions assuming the same values for $p_0$ and $n$ as for the $\pi^0$ fits, provide a good description of the data ($\chi^2$/d.f. = 10.1 / 5). A separate fit with 3 free parameters (dashed line) is almost identical in the range of the $\eta$ data ($\chi^2$/d.f. = 9.7 / 3). To further check the sensitivity of the data to $m_T$ scaling properties, we have analyzed the $\eta$ and $\pi^0$ invariant cross sections by fitting pure exponentials in the same $p_T$ and $m_T$ ranges, respectively. It turns out that in the $m_T$ representation the inverse slope parameter of the $\eta$ distribution, $T_{\eta,m_T} = (226 \pm 9)$ MeV, is only 4 % ($\cong 1\sigma$) larger, while in the $p_T$ representation it is 15 % ($\gtrsim 3\sigma$) larger than the corresponding values deduced from the $\pi^0$ spectra. These numbers demonstrate the sensitivity of the data and seem to indicate at most a very small deviation from $m_T$ scaling. In Ref. [10] the energy distribution of particles is interpreted as being composed of a thermal and a collective radial expansion component. Applying this picture to the observed (but not significant) deviation of the $\pi^0$ and $\eta$ $m_T$ slopes suggests a radial expansion velocity at freeze-out of appr. 30-50 % of the speed of light, well in line with the results of Ref. [10].

To test the concept of $m_T$-scaling in a model independent way, the differential $\eta/\pi^0$ ratio is presented in Fig. 4 for identical bins in $m_T$. The average of all data points yields a ratio of 0.53±0.07 for the S+Au data and 0.43±0.15 for the S+S data. Within the error bars and over a broad $m_T$ range we observe no significant deviation from a horizontal line, again demonstrating consistency with $m_T$ scaling.

Figure 5 shows as an alternative representation of the measured data the $\eta/\pi^0$-ratios as a function of $p_T$. The full symbols represent the results obtained within the present and previous experiment [14] for three different reaction systems (see also Table 1), whereas the open symbols show a compilation of various p+p, $\pi$+p, and p+C midrapidity [25, 26, 27, 28, 29, 31] and forward rapidity data [2, 3, 30]. The latter two have predominantly been taken at higher CM-energies ($\sqrt{s}$ = 19.4 – 62.0 GeV) as compared to the present experiment ($\sqrt{s}$ = 19.4 AGeV). A comparison of all data points might nevertheless be justified, since the $\eta/\pi^0$-ratios have been found to be insensitive to $\sqrt{s}$ in p-induced reactions [26]. For a quantitative comparison, the spectral shapes of the different data sets were taken into account again according to Eq. 1. Plotted as a function of $p_T$ the differential $\eta/\pi^0$-ratio is then written as

$$E\frac{d^3\sigma}{dp^3}(\eta)\Big/ E\frac{d^3\sigma}{dp^3}(\pi^0) = R_{\eta/\pi}\cdot\left(\frac{p_0+\sqrt{m_\pi^2+p_T^2}}{p_0+\sqrt{m_\eta^2+p_T^2}}\right)^n \tag{2}$$

with $p_0$ and $n$ as in Eq. 1 and $R_{\eta/\pi} = C_\eta/C_\pi$. Using $p_0$ and $n$ as extracted from the $\pi^0$ $m_T$ distributions, the differential $\eta/\pi^0$ cross section ratio is found to be $R^{S+Au}_{\eta/\pi} = 0.55 \pm 0.07$ in S+Au data (solid line in Fig. 5) and $R^{S+S}_{\eta/\pi} = 0.42 \pm 0.13$ in S+S data, well in agreement to the results above. For comparison, the differential $\eta/\pi^0$-ratio of the available p-induced data was extracted in the same way. Taking as an average $p_0 = 4.9 \pm 0.1$ GeV/$c$ and $n = 33.6 \pm 0.7$ from pion data of various experiments [25, 32, 33, 34, 35, 36] we find $R^{p}_{\eta/\pi} = 0.55 \pm 0.02$ (dashed line in Fig. 5).

It should be emphasized, that the numbers quoted above represent the cross section ratios measured for equal bins in $m_T$. Because $m_T \approx p_T$ for $p_T \gg m_0$, the same result would be obtained for large values of $p_T$. Extracting the *integrated* $\eta/\pi^0$ production ratio requires an extrapolation of the experimental $p_T$ spectra to $p_T \to 0$ , or likewise an extrapolation of the experimental $m_T$ spectra to $m_T \to m_\eta$ and $m_\pi$, respectively (see Eq. 2 and Fig. 5). Such an extrapolation into regions outside the experimental acceptance induces additional systematic uncertainties. Assuming the fit function of Eq. 1 the extrapolation yields a ratio of integrated production cross sections of $(\eta/\pi^0)_{S+Au} = 0.147 \pm 0.017(\text{stat.}) \pm 0.015(\text{syst.})$, $(\eta/\pi^0)_{S+S} = 0.120 \pm 0.034(\text{stat.}) \pm 0.022(\text{syst.})$, and $(\eta/\pi^0)_p = 0.103 \pm 0.003(\text{stat.}) \pm 0.006(\text{syst.})$ for the different reaction systems, respectively. As a consequence of the larger inverse slope of the spectra in S induced reactions, the integrated $\eta/\pi$ ratios appear to be enhanced over p induced reactions, even though the asymptotic ratios $R_{\eta/\pi}$ are very similar. For comparison with other mesons, the $K^+/\pi^+$ ratio in central 14.5 AGeV Si+Au collisions is measured to be $0.185 \pm 0.007$ [9], while a value of $0.137 \pm 0.008$ has been deduced from central 200 $A$GeV/$c$ S+S collisions [37]. Similarly, $K^0_s/\pi^- \cong 0.104 \pm 0.012$ may be inferred from 200 $A$GeV/$c$ S+Ag data [8]. Employing a simplified quantum statistical hadron-gas model in which relative particle production of sufficiently large systems is governed by their respective masses and in which modifications due to resonance decays are taken into account, a temperature larger than $T \gtrsim 150\,\text{MeV}$ is needed to reproduce the present experimental $\eta/\pi^0$ data [38].

| reaction system | $p_T$-range (GeV/$c$) | $\eta/\pi^0$-ratio | $m_T$-range (GeV/$c$) | $\eta/\pi^0$-ratio |
|---|---|---|---|---|
| S+Au | 0.5 − 0.7 | $0.175 \pm 0.066$ | 0.7 − 0.9 | $0.512 \pm 0.162$ |
| | 0.7 − 1.0 | $0.163 \pm 0.065$ | 0.9 − 1.1 | $0.250 \pm 0.149$ |
| | 1.0 − 1.5 | $0.423 \pm 0.087$ | 1.1 − 1.5 | $0.674 \pm 0.147$ |
| | 1.5 − 2.0 | $0.398 \pm 0.115$ | 1.5 − 2.0 | $0.461 \pm 0.148$ |
| | 2.0 − 2.5 | $0.705 \pm 0.183$ | 2.0 − 2.5 | $1.048 \pm 0.232$ |
| | 2.5 − 3.5 | $0.496 \pm 0.203$ | 2.5 − 3.5 | $0.581 \pm 0.231$ |
| S+S | 0.5 − 1.0 | $0.219 \pm 0.074$ | 0.7 − 1.1 | $0.510 \pm 0.194$ |
| | 1.0 − 1.5 | $0.173 \pm 0.122$ | 1.1 − 1.5 | $0.306 \pm 0.235$ |

Table 1: Measured $\eta/\pi^0$ cross section ratios for different bins in $p_T$ and $m_T$.

In summary, inclusive differential cross-sections of $\eta$ mesons have been measured for the first time in 200 $A$GeV/$c$ nuclear collisions. Together with the $\pi^0$ spectra measured within the same apparatus via the same decay mode, the results are compatible with the phenomenological concept of $m_T$ scaling. The slight deviation observed, can be well accounted for by assuming a radial expansion velocity of $\beta \approx$ 0.3-0.5. The ratio of normalizations of the $m_T$ spectra, or equivalently, the $\eta/\pi^0$-ratio at large $p_T$, were found to be $0.53 \pm 0.07$ for S+Au and $0.43 \pm 0.15$ for S+S collisions. Extrapolation to full phase space yields an integrated production cross section ratio of $0.147 \pm 0.017$(stat.)$\pm 0.015$(syst.), and $0.120 \pm 0.034$(stat.)$\pm$ 0.022(syst.), respectively, which is to be compared to $0.103 \pm 0.003$(stat.) $\pm$ 0.006(syst.) extracted from data of p-induced reactions.

This work was partially supported by the German BMBF and DFG, the United States DOE under contracts with BNL (DE-AC02-76CH00016), LBL (DE-AC03-76SF00098), and ORNL Lockheed Martin Energy Systems (DE-AC05-84OR21400), the Swedish NFR, the International Science Foundation (ISF), the Dutch Stiching FOM, and the CERN PPE division.

# References

[1] A summary of most recent data can be found in the proceedings to Quark-Matter conferences, QM95: Nucl. Phys. **A** (1995) and QM 93: Nucl. Phys. **A566** (1994).

[2] M. Aguilar-Benitez et al., (NA27 Collaboration), Z. Phys. **C50** (1991) 405.

[3] R. Apsimon et al., Z. Phys. **C54** (1992) 185.

[4] E. Shuryak and O. Zirhov, Phys. Lett. **B89** (1980) 253.

[5] R. Hagedorn, Riv. Nuovo Cimento **6** (1983) 1.

[6] Y. Tarasov, Sov. J. Nucl. Phys. **42** (1985) 260.

[7] H. Van Hecke et al., (NA34 Collaboration), Nucl. Phys. **A525** (1991) 227c.

[8] T. Alber et al., (NA35 Collaboration), Z. Phys. **C64** (1994) 195.

[9] T. Abbott et al., (E802 Collaboration), Phys. Rev **C50** (1994) 1024.

[10] K. Lee et al., Z. Phys. **C48** (1990) 425.

[11] H. Baumeister et al., Nucl. Instr. and Meth. **A292** (1990) 81.

[12] T. Awes et al., Nucl. Instr. and Meth. **A279** (1989) 479.

[13] R. Albrecht et al., (WA80 Collaboration), Z. Phys. **C47** (1990) 367.

[14] R. Albrecht et al., (WA80 Collaboration), Z. Phys. **C51** (1991) 1.

[15] F. Berger et al., Nucl. Instr. and Meth. **A321** (1992) 152.

[16] R. Albrecht et al., (WA80 Collaboration), Phys. Rev. **C50** (1994) 1048.

[17] T. Awes, Nucl. Instr. and Meth. **A276** (1989) 468.

[18] A. Lebedev et al., (WA80 Collaboration), Nucl. Phys. **A566** (1994) 355c.

[19] G. Hölker, Doctoral thesis, 1993, University of Münster.

[20] M. Abramowitz and I. Stegun, *Handbook of mathematical functions*, Dover publications Inc., New York, 1965.

[21] G. Clewing, Doctoral thesis, 1993, University of Münster.

[22] J. Harris et al., (NA35 Collaboration), Nucl. Phys. **A498** (1989) 133c.

[23] K. Guettler et al., Phys. Lett. **B64** (1976) 111.

[24] J. Bartke et al., Nucl. Phys. **B120** (1977) 14.

[25] F. Büsser et al., Nucl. Phys. **B106** (1976) 1.

[26] G. Donaldson et al., Phys. Rev. Lett. **40** (1978) 684.

[27] C. Kourkoumelis et al., Phys. Lett. **B84** (1979) 277.

[28] J. Povlis et al., Phys. Rev. Lett. **51** (1983) 967.

[29] T. Akesson et al., (AFS Collaboration), Phys. Lett. **B178** (1986) 447.

[30] J. Antille et al., (UA6 Collaboration), Phys. Lett. **B194** (1987) 568.

[31] M. Bonesini et al., (WA70 Collaboration), Z. Phys. **C42** (1989) 527.

[32] M. Adamus et al., (NA22 Collaboration), Z. Phys. **C39** (1988) 311.

[33] B. Alper et al., Nucl. Phys. **B100** (1975) 237.

[34] D. Antreasyan et al., Phys. Rev. **D19** (1979) 764.

[35] C. De Marzo et al., Phys. Rev. **D36** (1987) 16.

[36] G. Donaldson et al., Phys. Rev. Lett. **36** (1976) 1110.

[37] J. Bächler et al., (NA35 Collaboration), Z. Phys. **C58** (1993) 367.

[38] Peter Koch, private communication.

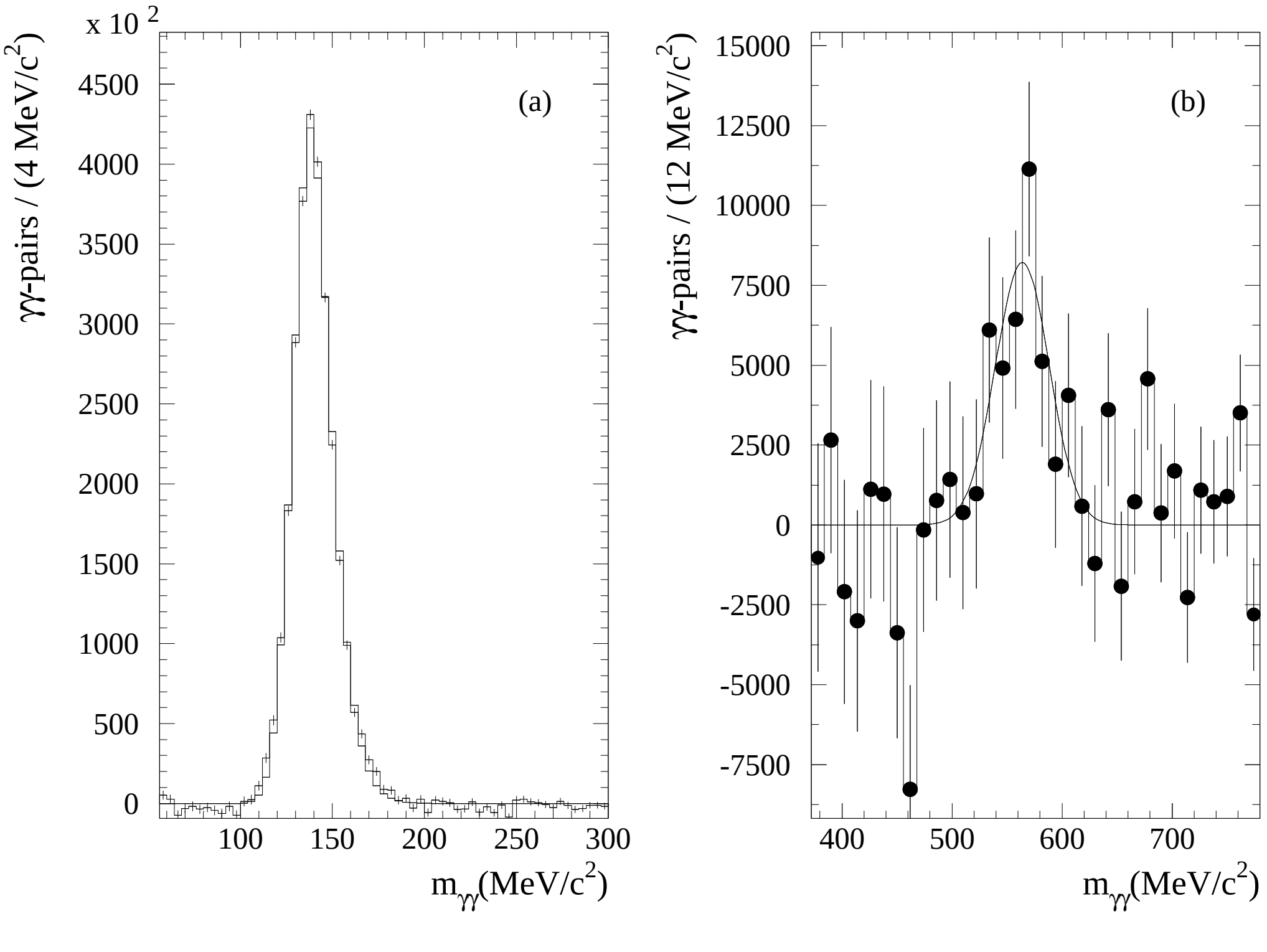

Figure 2: Background subtracted invariant mass distributions in the $\pi^0$ (a) and $\eta$ (b) mass range accumulated for transverse momenta of $0.5 \leq p_T \leq 1.0$ GeV/$c$. The signal to background ratio is 5.7 % for (a) and 0.07 % for (b).

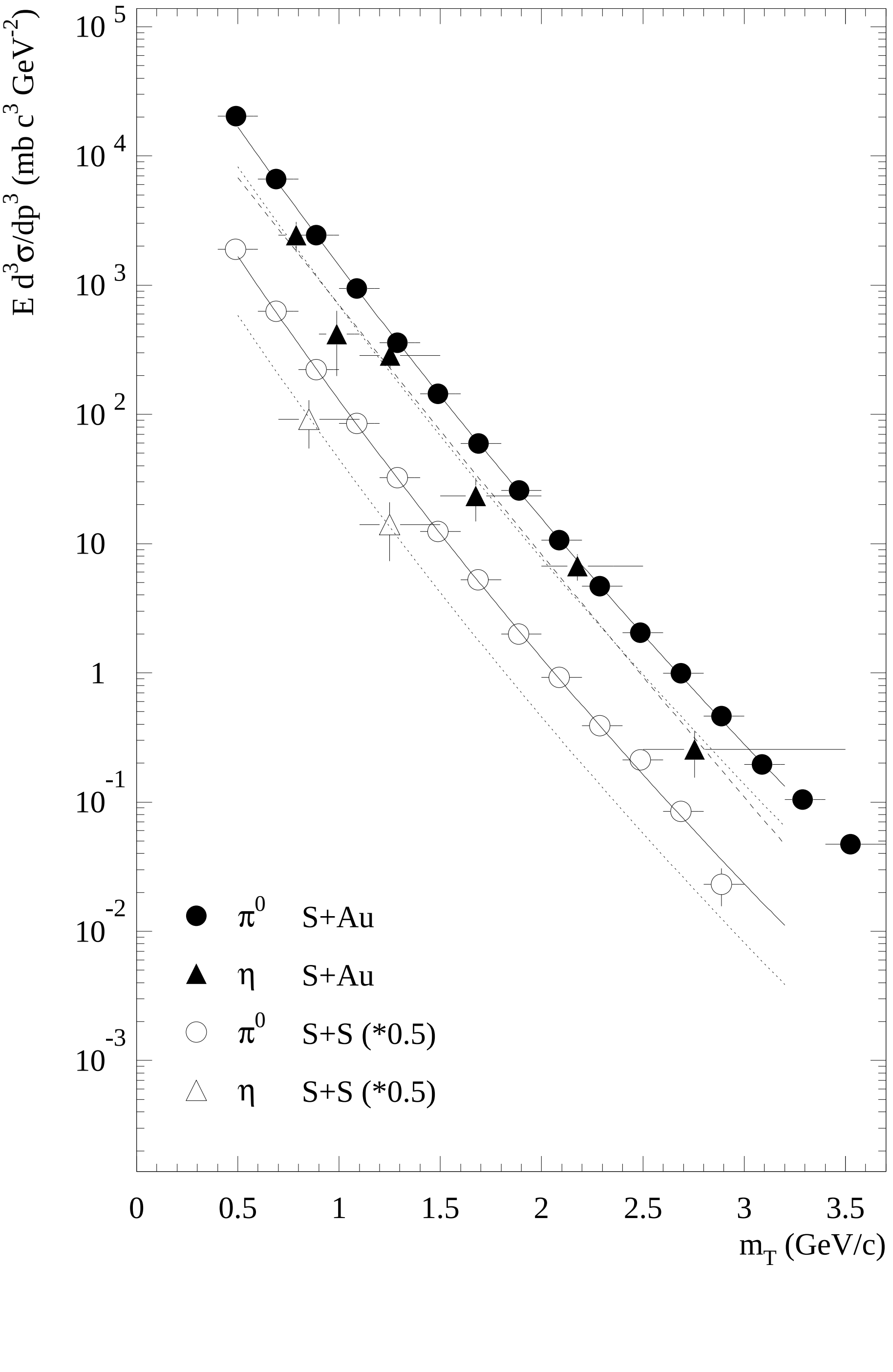

Figure 3: Invariant cross sections of $\pi^0$ and $\eta$ mesons averaged over the rapidity interval of $2.1 \leq y \leq 2.9$ as a function of transverse mass for 200 $A$GeV/$c$ S+Au and S+S minimum bias data. The S+S data have been scaled by a factor 0.5 for better presentation. The different lines are fitted results as explained in the text.

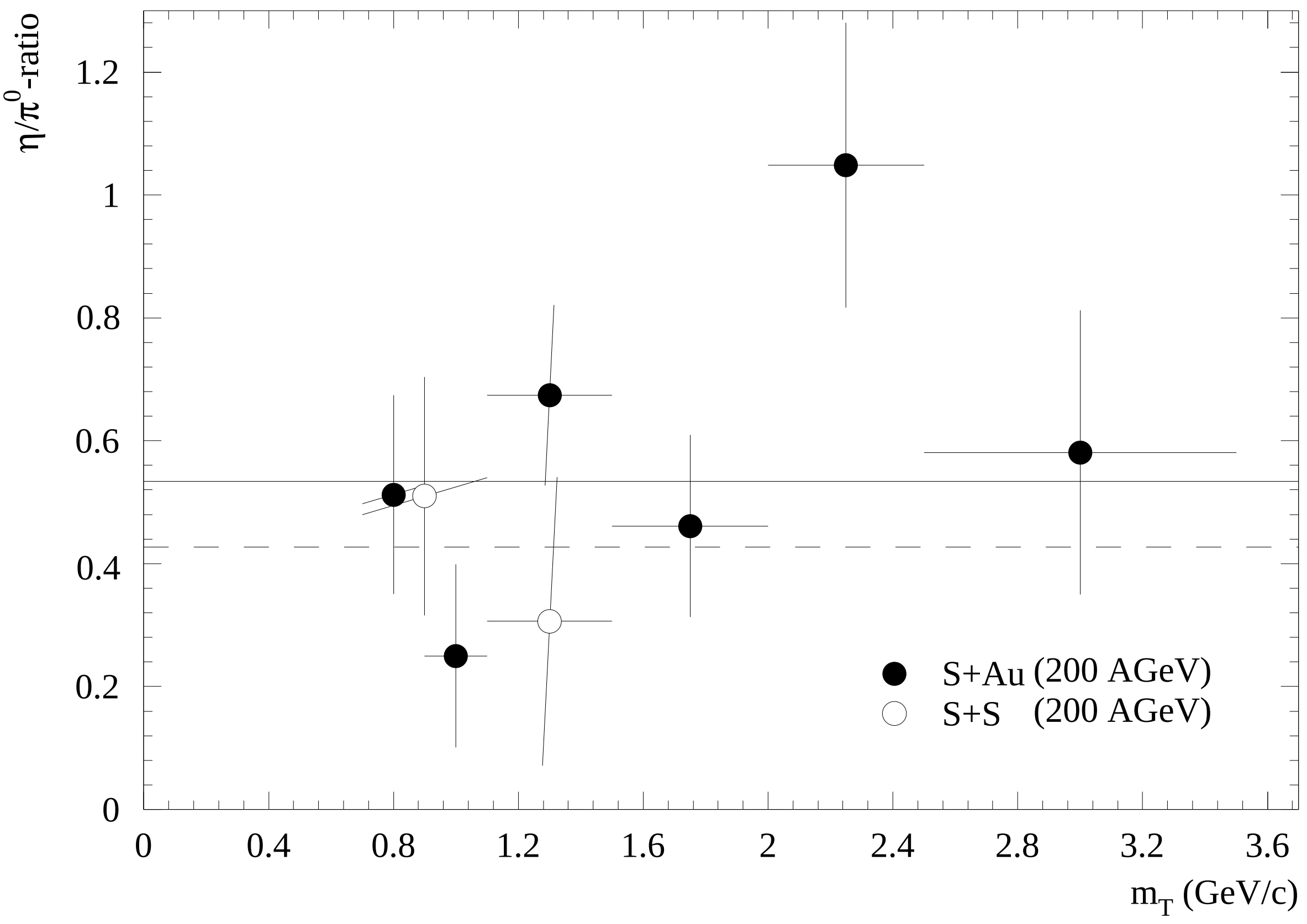

Figure 4: $\eta/\pi^0$-ratio as a function of $m_T$. The lines represent the mean values of the S+Au (solid line) and S+S data points (dashed line).

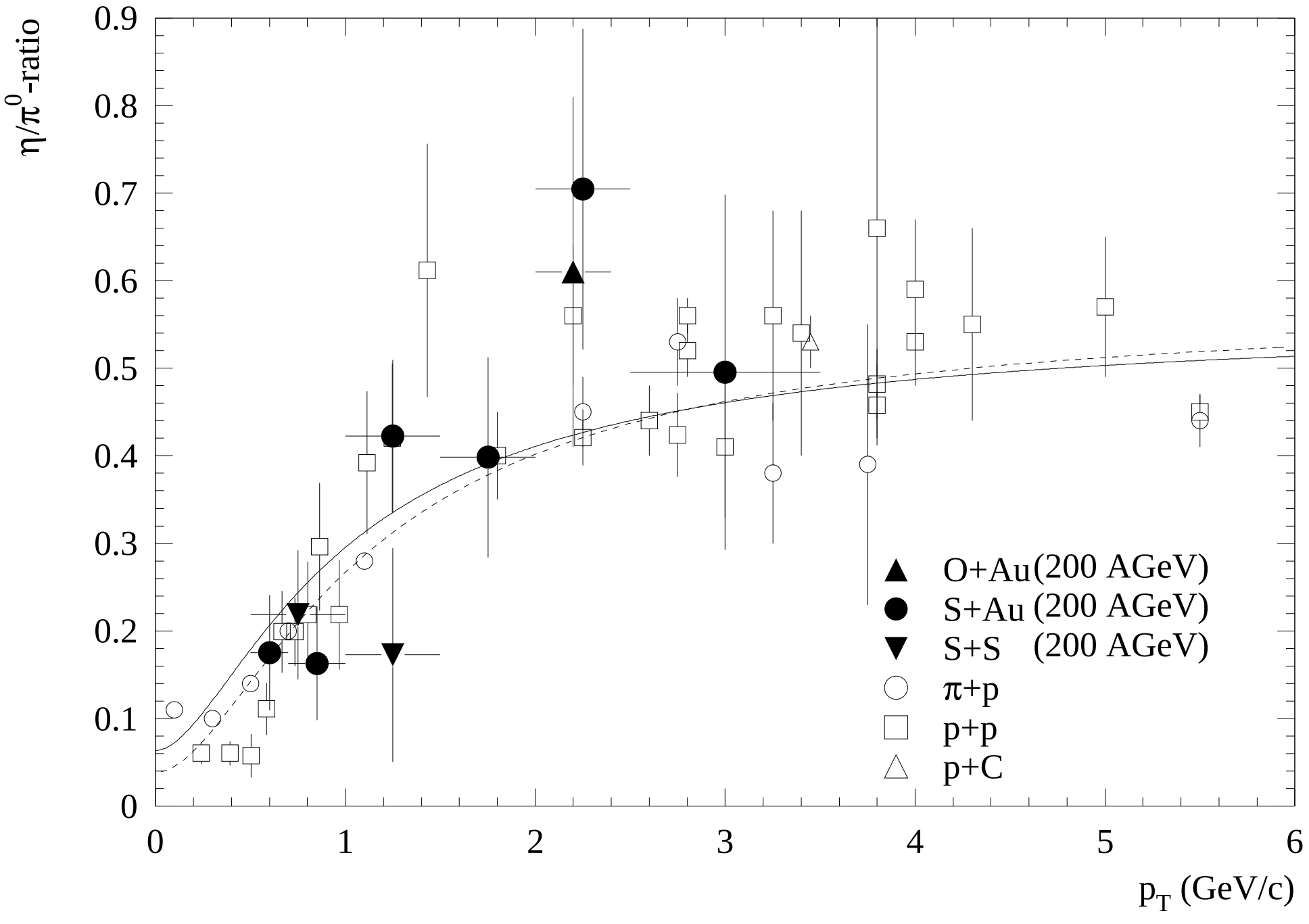

Figure 5: $\eta/\pi^0$-ratio as a function of $p_T$. The solid and dashed lines represent $m_T$ scaling parameterizations (Eq. 2) extracted from S+Au data and p induced collisions, respectively. (See text for details.)